%% file: main.tex
\documentclass[conference]{IEEEconf}

\input epsf
\usepackage{graphicx}
\usepackage{caption}
\usepackage{subcaption}
\usepackage{wrapfig}
\usepackage{multirow}
\usepackage{titlesec}
\usepackage{balance} 
\usepackage{stfloats}
\usepackage{xcolor}
\usepackage{listings}
\usepackage{hhline}
\usepackage{color}
\usepackage{booktabs}
\usepackage{multirow}
\usepackage{array}
\usepackage{blindtext}
\usepackage{cite}
\usepackage{url}

\renewcommand\thesection{\arabic{section}} 

\begin{document}

\title{\textbf{\Large Concolic Testing of JavaScript using Sparkplug\\}}

\author{Zhe Li$^{1}$, Fei Xie$^{2,*}$\\
	\normalsize $^{1}$$^{2}$Portland State University, Portland, OR, USA\\
	\normalsize zl3@pdx.edu, xie@pdx.edu\\
}

\maketitle

\begin{abstract}
\label{sec:abstract}
\input{00-abstract}
\end{abstract}

\section{Introduction}
\label{sec:intro}
\input{01-intro}

\section{Background}
\label{sec:background}
\input{02-background}

\section{Approach}
\label{sec:approach}
\input{03-approach}

\section{Implementations}
\label{sec:impl}
\input{04-implementation}

\section{Evaluations}
\label{sec:eval}
\input{05-evaluation}

\section{Related Work}
\label{sec:relatedwork}
\input{06-relatedwork}

\section{Conclusions}
\label{sec:conclusion}
\input{07-conclusion}

\bibliographystyle{IEEEtran}
\bibliography{nodejs}
\end{document}

%% file: 00-abstract.tex
JavaScript has become the most popular programming language for not only web front-end development but also a wide range of server-side applications. Many of such applications handle sensitive information such as financial transactions and private conversions. Errors in such applications not only affect user experiences, but also endanger the safety, security, and privacy of users. While the reliability of JavaScript code will be of more importance, testing techniques for the language remain insufficient, compared to other languages. In-situ concolic testing of JS scripts is a framework that enables concolic testing of JS scripts in their native environments and is able to automatically generate test cases. However, its \textit{Qemu} based execution tracing engine is slow in capturing traces, which is also not portable and involves two translation stages, which is a complex and error-prone process. In this paper, our approach proposed to deploy a new execution tracer leveraging V8's Sparkplug baseline compiler to improve the tracing process and a new assembly to LLVM IR using \textit{remill} libraries. We evaluated its effectiveness and efficiency by comparing the coverage, bug detection, and time consumption with the in-situ approach on the same test set, which are 160 Node.js libraries that heavily utilize the \texttt{String} type and its operations. The results show our approach achieves similar statement coverage on these libraries within no more than 10\% difference on average and is able to detect all bugs that are detected by the in-situ method and more, which only use a fraction of the time needed by the in-situ approach.

%% file: 01-intro.tex
Since its emergence as a scripting language for dynamic web elements, JavaScript (JS) has experienced a surge in popularity and has evolved into a versatile and extensively utilized application programming language. The Node.js runtime, leveraging Chrome's V8 JS engine as its foundation, empowers developers to create a diverse range of server-side and client-side browser-less applications using pure JavaScript~\cite{nodejs}. An entire ecosystem of Node.js libraries has been cultivated, accessible via the Node Package Manager (NPM), and extensively employed in the development of applications~\cite{npm}. As JavaScript continues to gain significance in the web, mobile, and cloud infrastructure of modern systems, the repercussions of bugs and security vulnerabilities in JS scripts become increasingly severe~\cite{npmsecurity}. JS scripts, whether browser or Node.js based, are often perceived by many developers as significant security vulnerabilities. Common security concerns associated with browser-based JS scripts include cross-site scripting (XSS) and SQL injection (SQLi)~\cite{jssecurity}, etc. Errors and failures in JS scripts executed on Node.js can result in server crashes or compromises. Among the most prevalent security issues in Node.js are NPM phishing and denial of service (DoS) attacks targeting regular expressions. NPM provides developers with the capability to develop and distribute JS libraries for reuse, yet this flexibility introduces notable security risks~\cite{van2019server}. Developers face a pressing demand to construct comprehensive test suites capable of early bug and security vulnerability detection. However, manually crafting such suites has become a costly and time-consuming bottleneck in software development~\cite{mirshokraie2015jseft}.

Symbolic execution is a powerful technique for automating the generation of test cases and identifying bugs in real-world software. It entails executing a program using symbolic values, monitoring program path conditions via symbolic expressions, and producing test cases to explore these paths by solving symbolic path conditions~\cite{king1976symbolic}. Concolic testing is a hybrid verification technique designed to address the challenge of path explosion often encountered in symbolic execution~\cite{krishnamoorthy2010tackling}. Concolic testing employs symbolic execution to traverse only the branches along a concrete execution path determined by a concrete input of the program being tested. This approach effectively reduces the explored path space, mitigating the issue of path explosion~\cite{sen2007concolic}. Traditional symbolic or concolic execution engines primarily focus on analyzing code written in languages like C/C++ or those that compile to low-level intermediate representations (LLVM)~\cite{llvm} or binary code, e.g., KLEE~\cite{klee}, BitBlaze~\cite{bitblaze}, S2E~\cite{s2e}, DART~\cite{dart}, CUTE~\cite{cute}, SAGE~\cite{sage}, and CRETE~\cite{crete}.

In-situ concolic testing of JS scripts is a novel framework that enables concolic testing of JS scripts in their native environments and can automatically generate test cases that achieve comparable, if not better, code coverage than manually crafted unit test suites for Node.js libraries and discovered previously unknown bugs in these libraries~\cite{insitu-js}. Most approaches of concolic testing on JavaScript typically take JS scripts out of their native execution environments and analyze them in artificial test harnesses. For example, the Kudzu engine addresses the problem of client-side code injection vulnerabilities for JavaScript~\cite{kudzu}. It involves modifying the JS interpreter to build a new symbolic execution engine, which requires significant effort in implementation and maintenance. Such JS-specific symbolic engines have not demonstrated the effectiveness and efficiency that warrants wide adoption~\cite{spejs}. In-situ concolic testing for JavaScript using JavaScript's native execution environments becomes its biggest strength. However, it has several limitations~\cite{insitu-js}. It utilized the tracing engine of CRETE, which leverages the interpreted mode of \textit{Qemu}, a dynamic translator~\cite{bellard2005qemu}, to capture the execution trace of JS scripts and uses \texttt{KLEE} as the backend symbolic execution engine. The concrete execution trace is converted from a piece of code to the host instruction set, and the instruction set is then translated to \texttt{qemu-ir} by the tiny code generator (TCG) of \textit{Qemu} dynamic translation backend. This process hinders the efficiency of the tracing process greatly since the in-situ approach uses the interpreted mode with TCG to enable tracing. The execution tracer of CRETE takes 3 minutes to trace a JS function with 12 lines of code on average, which is inefficient. The execution traces are then translated from \texttt{qemu-ir} to \texttt{LLVM IR} by an offline translator based on \texttt{$S^2E$}. This workflow involves two stages of translation for the execution traces, which gives more chances for introducing errors and mistakes. 

To improve the efficiency of the execution tracer, reduce the number of translation stages, and conduct concolic testing in their native environments like the in-situ approach at the same time, our approach proposed to deploy a new execution tracer leveraging V8's Sparkplug baseline compiler to improve the tracing process and a new assembly to LLVM IR using \textit{remill} libraries in this paper. We evaluated its effectiveness and efficiency by comparing the coverage, bug detection, and time consumption with the in-situ approach on the same test set, which are 160 Node.js libraries that heavily utilize the \texttt{String} type and its operations. The results show our approach achieves similar statement coverage on these libraries within no more than 10\% difference on average and is able to detect all bugs that are found by the in-situ method, which only uses a fraction of the time needed by the in-situ approach.

%% file: 02-background.tex
\subsection{Sparkplug}
\label{subsec:sparkplug}
\begin{figure}
    \centering
    \includegraphics[width=\columnwidth]{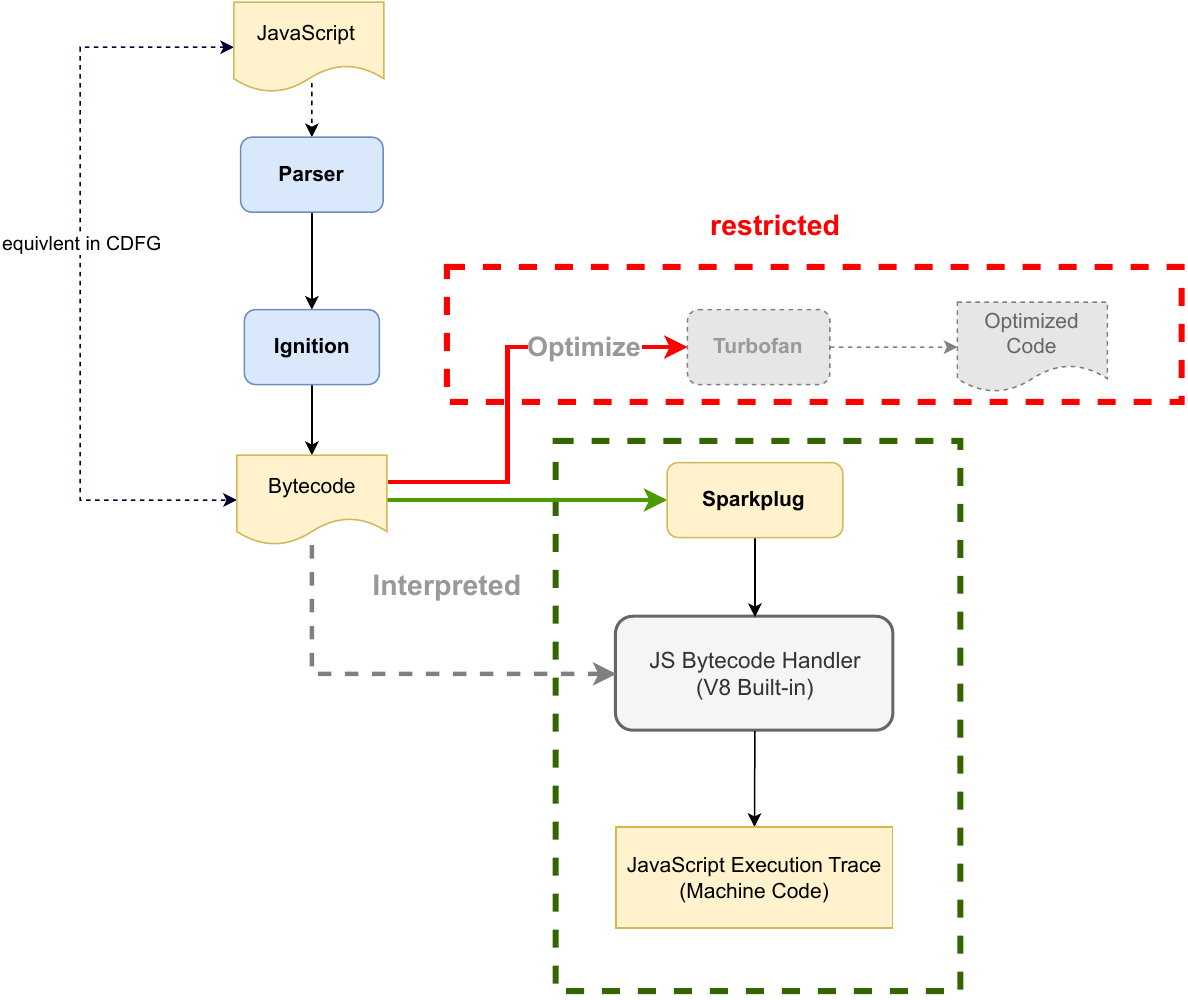}
    \caption{Sparkplug Restricted optimization feature}
    \label{fig:sparchbenifit}
\end{figure}

Sparkplug is a non-optimizing JavaScript compiler of V8~\cite{sparkplug}. It is engineered for swift compilation. Its speed is remarkable, enabling us to compile at our convenience, thereby facilitating a more aggressive tiering up to Sparkplug code~\cite{sparkplug}. There are a couple of techniques employed by the Sparkplug compiler to achieve its impressive speed. Firstly, it utilizes a shortcut; the functions it compiles are already processed into bytecode by a prior stage, which handles complex tasks such as variable resolution and parsing arrow functions. Sparkplug bypasses these intricate processes by compiling JavaScript from bytecode rather than directly from source code. Secondly, Sparkplug adopts a unique approach by skipping the generation of an intermediate representation (IR), a step typical in most compilers. Instead, it directly translates bytecode into machine code in a single linear pass using bytecode handlers~\cite{sparkplug}, aligning the emitted code with the execution flow of the bytecode. We will discuss the bytecode handler in detail in Section~\ref{sec:impl}. This feature guarantees that the emitted machine code execution trace we used for concolic analysis represents the execution flow of the source code. Remarkably, the entire Sparkplug compiler operates within a switch statement nested within a \texttt{for} loop, efficiently dispatching to predetermined bytecode handlers, the machine code generation functions based on the bytecode encountered. The absence of an IR restricts optimization opportunities to localized peephole optimizations as shown in Figure~\ref{fig:sparchbenifit}, we heavily utilize this feature of Sparkplug.

\subsection{Interpreter Stack Frame Mirroring}
\label{subsec:frames}
V8 JavaScript engine supports two modes for executing a JS script, namely \textit{interpreted mode} and \textit{optimized just-in-time compilation mode}. The \textit{interpreted mode} is where the JS bytecode~\cite{jsbytecode} translated from the JS script is interpreted by its interpreter, Ignition~\cite{ignition}, which is the foundation of in-situ concolic execution for JavaScript~\cite{insitu}. The optimized \textit{just-in-time compilation mode} is where the bytecode is compiled by the V8 engine into optimized machine code using its just-in-time compiler, Turbofan~\cite{v8unifiedcodegen}, and then executed on the target machine. Sparkplug as a baseline JavaScript compiler can restrict JS script from being optimized to mitigate complexity for later concolic execution. Furthermore, Sparkplug mirrors the execution of Ignition for JavaScript. Sparkplug intentionally aligns its stack frame layout with that of Ignition, ensuring that when Ignition stores a value in a register, Sparkplug does the same. This design choice simplifies Sparkplug compilation by allowing it to mirror the behavior of Ignition without the need for complex mappings between interpreter registers and Sparkplug's state. Therefore, it also allow us to improve the efficiency for the in-situ approach and keep its effectiveness at the same time. Sparkplug primarily consists of bytecode handler calls, which are short sequences of machine code embedded within the binary, along with control flow. Ignition and Sparkplug share significant portions of the bytecode handlers. In essence, Sparkplug serves as a serialization of Ignition execution, invoking the same built-ins and maintaining identical stack frames. This feature allows us to trace JS bytecode execution in its corresponding machine code like the execution in Ignition. Futhermore, Sparkplug effectively pre-compiles certain unavoidable interpreter overheads, such as operand decoding and dispatching to the next bytecode. This streamlined strategy contributes to Sparkplug's efficiency and performance. Therefore, Sparkplug can generate machine code that contains the same control flow as JS script, which can be used for code translation from machine code (assembly code) to LLVM.

\subsection{Remill}
McSema is an executable lifter that specializes in converting executable binaries from their machine code into LLVM. This process enables the translation of low-level binary instructions into a higher-level intermediate representation. Within McSema~\cite{mcsema}, the instruction translation functionality is powered by the \textit{Remill} library. Unlike some other tools, \textit{Remill} exclusively handles machine code translation into LLVM~\cite{remill}.

The versatility of \textit{Remill} extends to both static and dynamic binary translation scenarios. Notably, it has been employed in symbolic execution workflows alongside tools like KLEE~\cite{klee}. KLEE, which performs symbolic execution, typically operates on the LLVM IR generated from source code using the LLVM toolchain~\cite{llvm}. By utilizing \textit{Remill} to machine code into the LLVM IR, previously inaccessible targets become available for analysis with KLEE, thus expanding the range of symbolic execution capabilities.

\textit{Remill} delegates the implementation of memory accesses and specific types of control flow to the consumers of the generated LLVM. This deferral is facilitated through \textit{Remill} intrinsics, which are special functions representing various actions within the translated program. For instance, the \texttt{\_\_remill\_read\_memory} intrinsic function symbolizes the act of reading 8 bits of memory. By leveraging these intrinsics, downstream tools can differentiate between LLVM load and store instructions and access to the modeled program's memory. Moreover, downstream tools have the flexibility to implement memory intrinsics using LLVM's native memory access instructions. This approach allows us to create a seamless integration of \textit{Remill} generated LLVM into existing LLVM-based workflows while providing the necessary flexibility for custom memory access implementations tailored to specific analysis requirements. We utilized this feature to adapt the output to LLVM-based symbolic analysis tools.

%% file: 03-approach.tex
\subsection{Overview of goals}
\begin{figure*}
    \includegraphics[width=\textwidth]{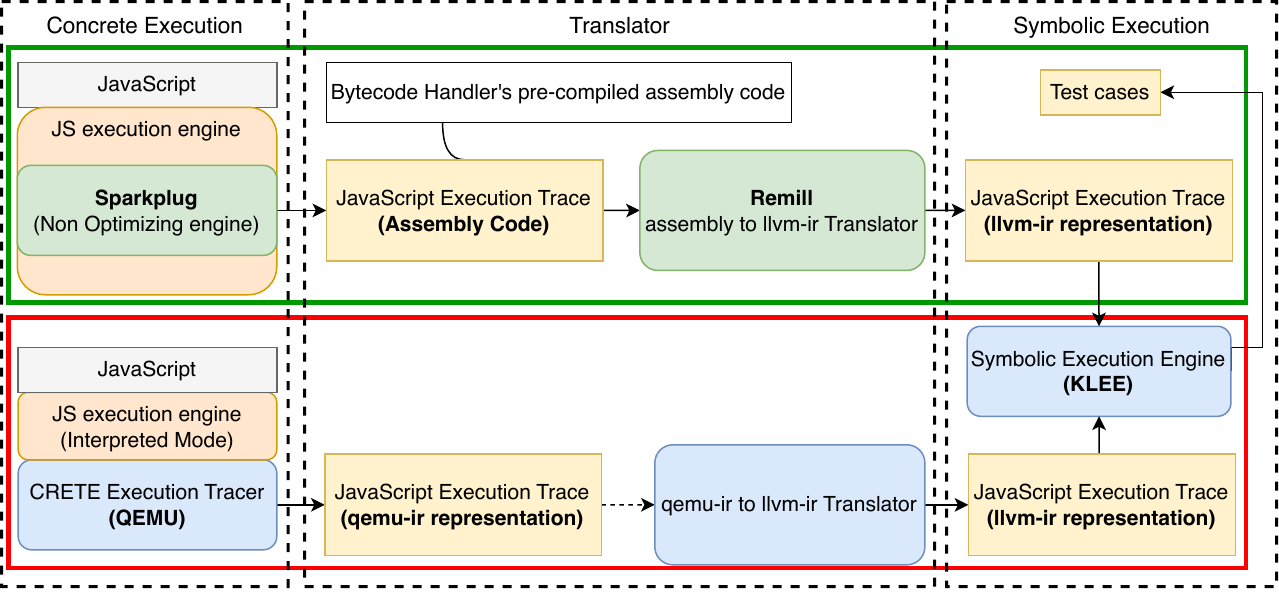}
    \caption{Workflows of In-situ Concolic Testing Based on Sparkplug and CRETE}
    \label{fig:workflowcompare}
\end{figure*}

Our approach aims to make improvements in efficiency for the in-situ approach mainly in generating execution traces and execution trace translation. Our approach strives to apply concolic testing on JS scripts in their native environment to generate effective test data for unit testing of these scripts. The workflow of concolic execution on JS scripts contains the following steps. As shown in Figure~\ref{fig:workflowcompare}, the concrete execution step in the leftmost box of concolic testing is conducted in the native execution environment for JS scripts, where the trace of this concrete execution is captured using the JS execution tracer. The trace is then analyzed in the symbolic execution step in the rightmost box of concolic testing to generate test cases automatically.
\begin{itemize}
    \item{\bf Execution trace capture.} Concrete execution traces of JS scripts are captured with a JS execution tracer, which is the interpretation of JavaScript bytecode. The concrete execution traces are in the form of assembly code, which represents the interpretation of JS bytecode execution.
    \item{\bf Translation.} In this step, our approach uses a translator to translate assembly code generated by the JS execution tracer into LLVM bitcode. 
    \item{\bf Symbolic Analysis.} The execution trace represented by LLVM is fed into a symbolic execution engine to generate test cases.
\end{itemize}

\subsection{Improvement}
In-situ concolic testing offers the capability of tracing inside the V8 JS engine to capture the execution trace that closely matches the JS bytecode interpretation~\cite{insitu-js,jsbytecode}. The conciseness of an execution trace determines the efficiency and effectiveness of later symbolic analysis and test case generation. Therefore, we intend to preserve such traits and achieve improvement of execution efficiency at the same time. Our approach improves in-situ concolic testing in 2 aspects. In Figure~\ref{fig:workflowcompare}, the in-situ approach is represented by the diagram in the red box and our approach in the green box. Firstly, compared to the in-situ approach of concolic testing for scripting languages, our approach frees the execution tracer from dependence on an emulator, which is normally slow. The concrete execution is obtained by the execution tracer, which leverages V8's Sparkplug engine instead of \texttt{CRETE} execution tracer based on \textit{qemu} in the in-situ approach. This speeds up the execution trace capture process since \textit{qemu} is based on an emulator. At the same time, it preserves the character that the execution trace capture happens in the native execution environment for JS script because we leverage the native Sparkplug baseline engine as the execution tracer. 

\begin{figure}
    \includegraphics[width=\columnwidth]{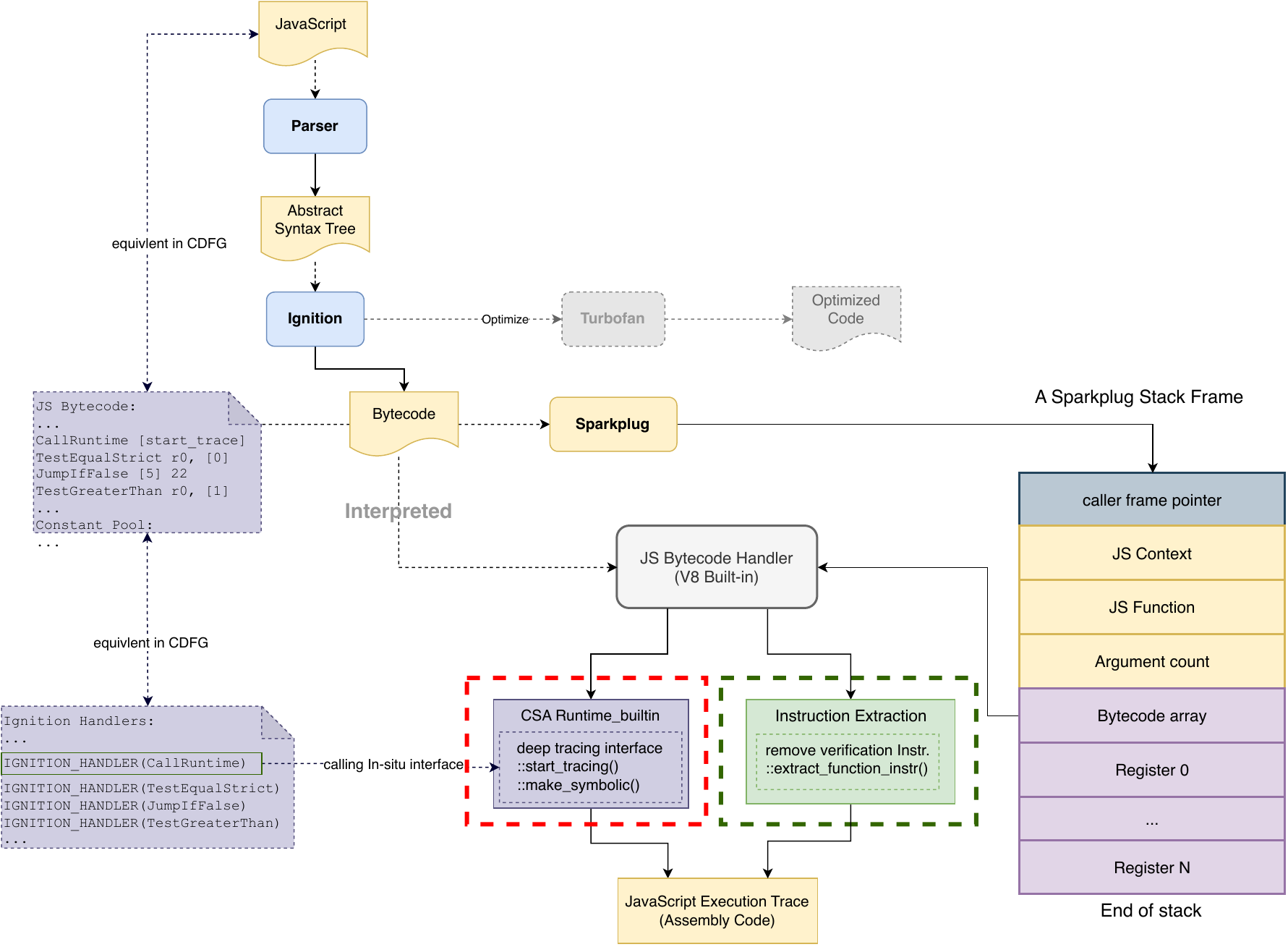}
    \caption{Workflow of Execution Tracer between In-situ Approach and Our Approach}
    \label{fig:executiontracer}
\end{figure}

\paragraph{Why we choose Sparkplug?}
Sparkplug disables the Turbofan path naturally. It compiles from bytecodes that Ignition emits as shown in Figure~\ref{fig:executiontracer}. JS bytecode preserves all necessary control flow JS source code has. Therefore, execution traces captured by Sparkplug have a one-to-one correspondence to the JS source code. The execution tracer based on Sparkplug directly traces the bytecodes translated from JS source code inside of V8. Furthermore, as mentioned in Section~\ref{subsec:frames}, Sparkplug mirrors Ignition's execution for JavaScript, Sparkplug and Ignition have almost identical stack frame~\cite{sparkplug}. This simplifies the design by removing the deep tracing control interface used in the in-situ approach shown in the red box by actually tracing within Sparkplug. To retrieve the most concise execution trace for JS script, our approach only extracts bytecodes that contribute to the control flow of JS script execution with \textit{Instruction Extraction} component, which removes the stack verification-related bytecodes in the generated execution trace without influencing the verification workflow of Sparkplug. 

\begin{figure}
    \includegraphics[width=\columnwidth]{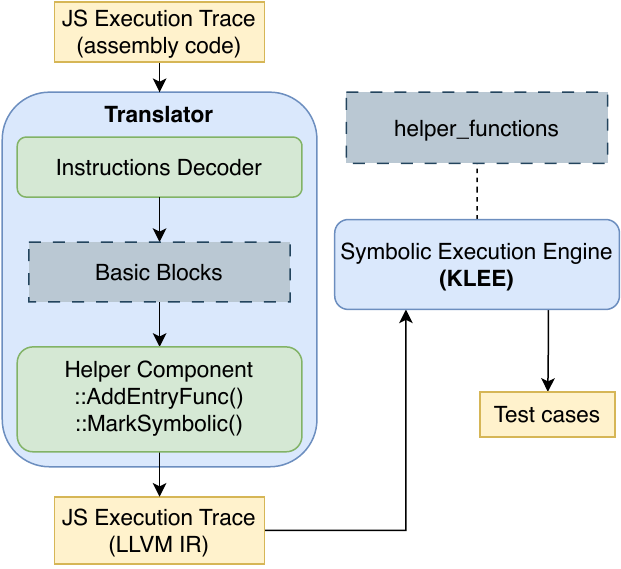}
    \caption{Workflow of the Translator}
    \label{fig:translator}
\end{figure}

Secondly, the in-situ approach uses an offline translator to translate \texttt{qemu-ir} to \texttt{LLVM IR}. \textit{Qemu}, the emulator first translates assembly code to the intermediate presentation of \texttt{qemu-ir} and then uses an offline translator to translate \texttt{qemu-ir} to \texttt{LLVM IR}. LLVM is a widely used intermediate presentation for symbolic analysis. Our approach simplifies this process by directly translating the captured execution traces from assembly code to \texttt{LLVM IR} shown in the middle box in Figure~\ref{fig:workflowcompare}. In this process, we introduce a helper component in the translator. This helper component aims to make the translated execution trace amenable to symbolic analysis tools by providing the main entry point and marking symbolic variables as shown in Figure~\ref{fig:translator}. As a result, the output of the translator produces a complete concrete execution trace for later symbolic execution engine to generate test cases.

%% file: 04-implementation.tex
In this section, we demonstrate the feasibility of our approach by implementing its complete workflow with an execution tracer based on V8's Sparkplug, a translator leveraging \texttt{Remill}, and a symbolic execution engine using \texttt{KLEE}~\cite{klee}. 

\begin{figure*}[htbp]
\centering
\includegraphics[width=\textwidth]{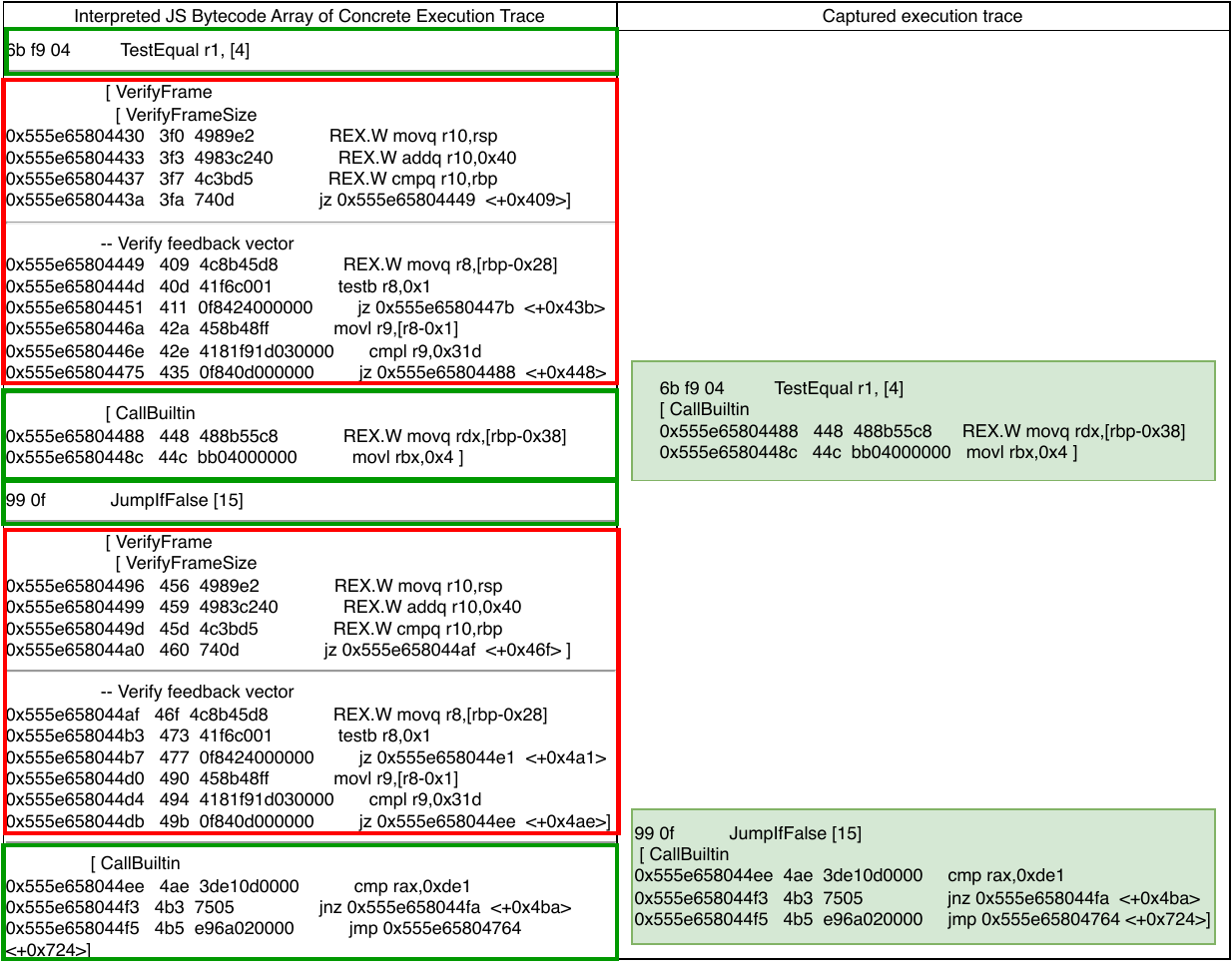}
\caption{How the execution tracer only extracts the execution traces that contribute to the main control flow of JS scripts}
\label{fig:tracerimpl}
\end{figure*}

\paragraph{Modification on Bytecode Handlers of Sparkplug}
To capture the most concise execution trace, we implemented the function \texttt{extract\_function\_instr} to filter out the stack verification-related compilation from Sparkplug and only extract the execution trace for bytecodes that contribute to the control flow of JS scripts. The left column of Figure~\ref{fig:tracerimpl} shows an example of an interpreted JS bytecode array of a concrete execution trace. Before interpreting each bytecode, Sparkplug verifies frame size and feedback vector. The execution tracer based on Sparkplug only removes the corresponding interpretation from the execution trace without changing Sparkplug's behavior. The green box indicates the bytecode extracted by the function and its correspondence assembly code generated by the bytecode handler of Sparkplug. The red box indicates the assembly instructions that are filtered out, which corresponds to stack frame verification. A special bytecode handler \texttt{PIN\_SYMBOLIC} is implemented to cache the symbolic value in the execution trace for later symbolic analysis.

\begin{figure}[htbp]
\centering
\includegraphics[width=\columnwidth]{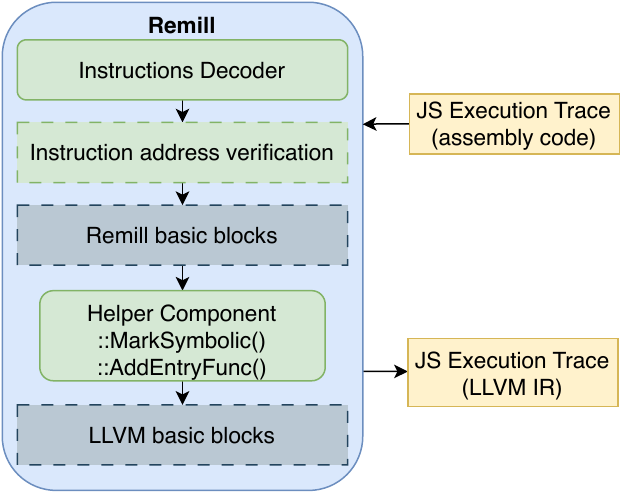}
\caption{How the execution tracer only extracts the execution traces that contribute to the main control flow of JS scripts}
\label{fig:translatorimpl}
\end{figure}

\paragraph{Implementation on Remill translator}
We utilized \textit{remill} library to implement an assembly-to-LLVM translator. Figure~\ref{fig:translatorimpl} shows the important components we implemented for the translator. It first checks if an instruction is valid as in whether the memory is executable and readable. In this process, it identifies the symbolic memory we cached by the execution tracer based on Sparkplug. After the correctness check, the translator translates \textit{remill} basic blocks to LLVM basic blocks. A helper component is added to create a main entry function to make the trace a self-contained LLVM module and mark symbolic memory for later symbolic analysis. The main function then calls into the basic blocks LLVM functions. At last, the resulting trace is readily consumable by the \texttt{KLEE}. We tested the execution tracer to ensure its correctness on 16 combinations of instructions such as math functions, basic arithmetic, \texttt{for} loop, \texttt{if-else}, etc.

During the symbolic execution stage, \texttt{KLEE} is modified to recognize the \textit{remill} intrinsic function for log error and exception. The execution trace is fed into \texttt{KLEE} to generate test cases. Test cases are used as a seed for the next iteration of symbolic execution to generate a comprehensive set of test cases, which is done by execution harness scripts.

%% file: 05-evaluation.tex
For evaluations, we targeted 160 Node.js libraries used in the in-situ approach to show the effectiveness and efficiencies after improvement. For effectiveness, we calculated the average time used for executing all libraries between the two methods. For efficiency, we evaluated the code coverage achieved by two methods. This evaluation is carried out on a Ubuntu OS Version 18.04 with 4-core Intel(R) Core(TM) i7-4790 CPU @ 3.60GHz and 16G memory.

To compare the two methods with these libraries, we built a test harness to systematically exercise all exported (public) methods in a given library with arguments whose type is \texttt{String}. The seed test cases are generated randomly within the test harness. We implemented an automation pipeline that helps set up the concolic testing environment for each Node.js library automatically. Coverage for all libraries is calculated using \textit{istanbul}, a popular JS coverage tool used by V8~\cite{istanbul} and compatible with most JavaScript testing frameworks, e.g., Mocha~\cite{mocha} and Node-Tap~\cite{tap}. Coverage may vary slightly due to the randomness of the seed test case generation. By default, the coverage that we show in this evaluation is statement coverage. 
\begin{table}[htb]
  \small
  \centering
  \vspace{-0.1in}
  \begin{tabular}{|l|l|l|}
    \hline
    \textbf{Metric} & \textbf{Range} & \textbf{Average}\\
    \hline
    Line of Code & [93, 16910] & 1687 \\
    \hline
    Weekly Downloads & [3, 37491350] & 9552965 \\ 
    \hline
    Dependencies & [3, 18154] & 282 \\
    \hline
  \end{tabular}\\
  \caption{Demographics for Libraries under Test}
  \label{tab:demographicsoflibs}
\vspace{-0.1in}
\end{table}
Table~\ref{tab:demographicsoflibs} shows the demographics of the selected libraries. The LoC (lines of code) for a library under test is calculated with \textit{github-loc}~\cite{githubloc}. The number of weekly downloads of a library under test is calculated with \textit{npm-stats-api}~\cite{npmstatsapi}. The number of dependencies is the number of dependent libraries that the library under test has. We calculated it with \textit{dependent-counts}~\cite{dependentcounts}.

\subsection{Coverage Analysis}
\begin{figure}[htbp]
\centering
\includegraphics[width=\columnwidth]{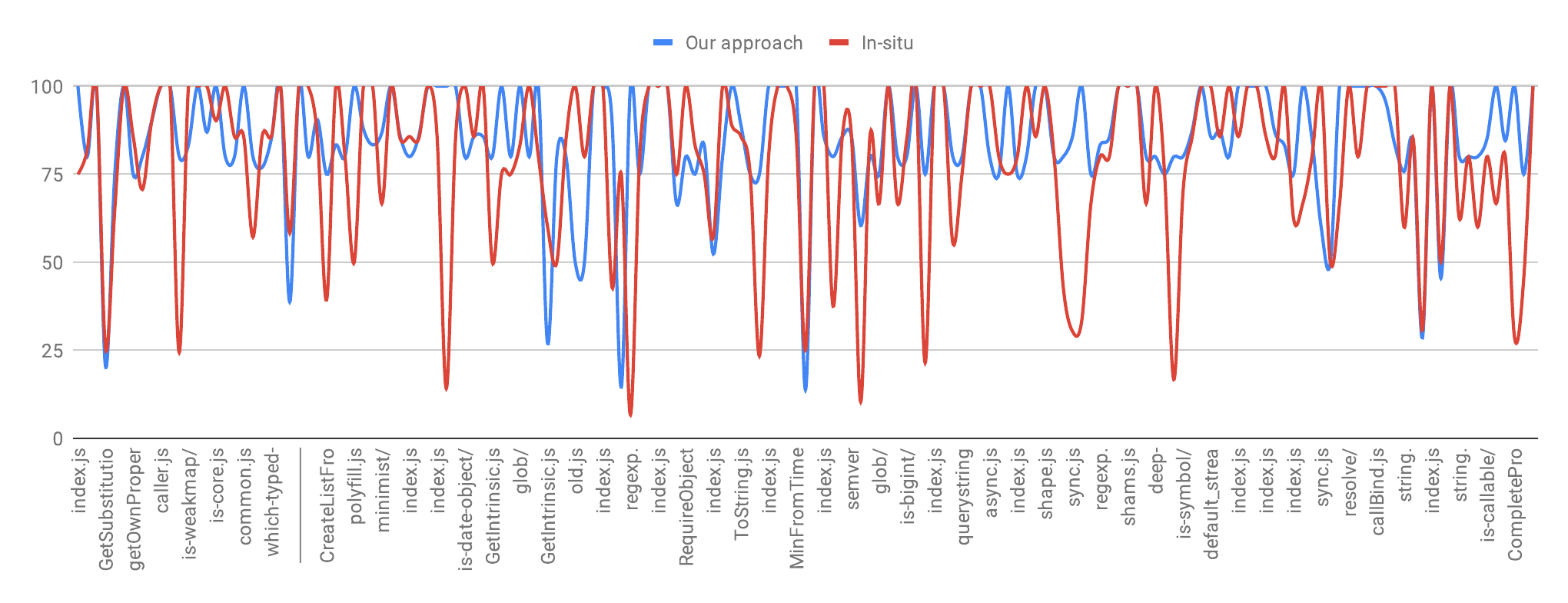}
\caption{Statement Coverage Comparison between our approach and In-situ approach}
\label{fig:coveragecompare}
\end{figure}

\begin{figure}[htbp]
\centering
\includegraphics[width=\columnwidth]{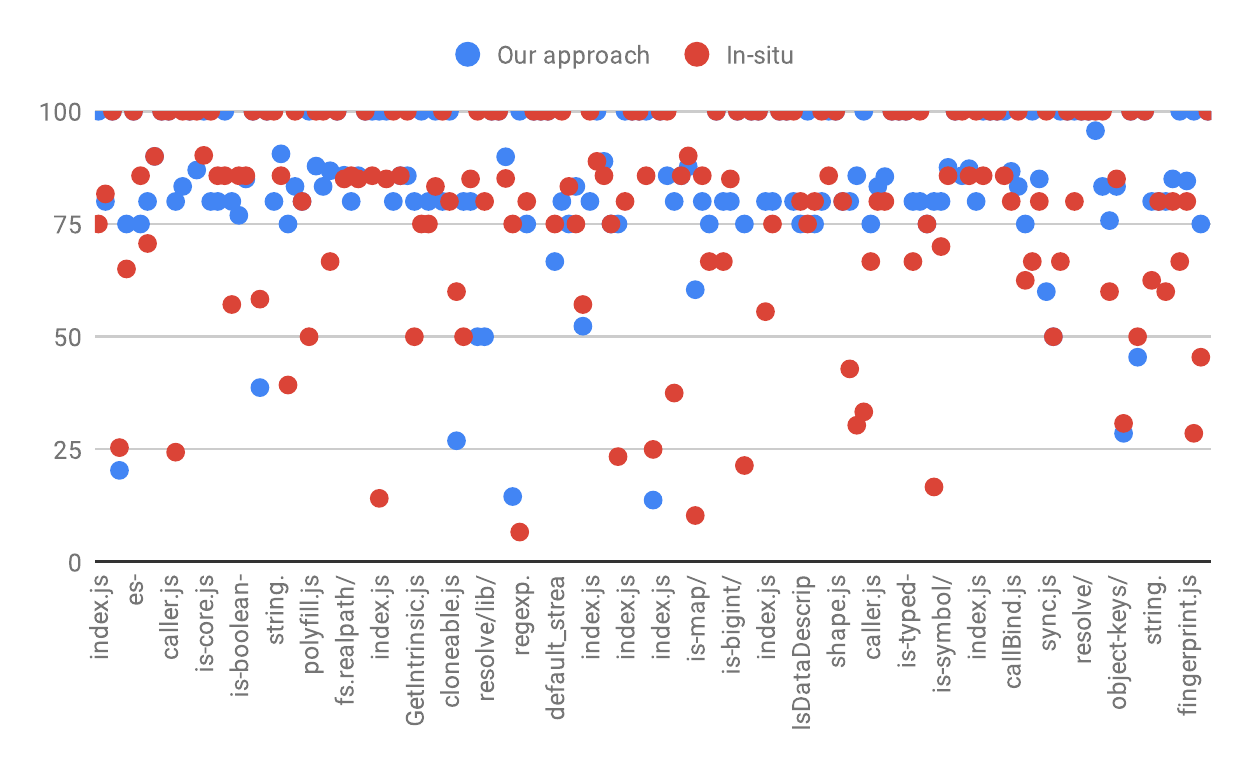}
\caption{Coverage Distribution Comparison between our approach and In-situ approach}
\label{fig:coveragecomparedots}
\end{figure}

Figure~\ref{fig:coveragecompare} shows the comparison of statement coverage achieved between our approach and the in-situ approach. The red line presented the statement coverage of the in-situ approach and the blue line indicates the statement coverage of our approach of improvement. We can see that they represent a similar trend of achieving statement coverage over 160 Node.js libraries under test. Figure~\ref{fig:coveragecomparedots} indicates the distribution of statement coverage between the two approaches, where the red dots represent the result of the in-situ approach and the blue dots indicate that of our approach. We can see major dots of both colors fall above the line of coverage of 75\%. Only 9 libraries achieved a coverage below 25\% and the reason is that it is a function with multiple arguments of \texttt{String} type, which can be made symbolic. Our test harness did not catch all of the arguments and only managed to set one of them as symbolic input. Therefore, it only explored the branches that are related to that one argument we set as symbolic input within the test harness. Among the libraries achieved below the coverage of 75\%, the red dots appear more times than the blue dots, which indicates our approach achieved higher coverage on average.

We also compared our approach with an existing tool, ExpoSE~\cite{loring2017expose}, by testing the same set of libraries as shown in Figure~\ref{fig:comparisonwithrelatedworks}, on which ExpoSE has been applied. Our method and the in-situ approach achieved similar higher coverage consistently. This comparison only partially reflects our method's ability to achieve higher coverage since ExpoSE mainly targets solving regular expression problems for its symbolic execution engine JALANGI.

\begin{figure}[htbp]
\centering
\includegraphics[width=\columnwidth]{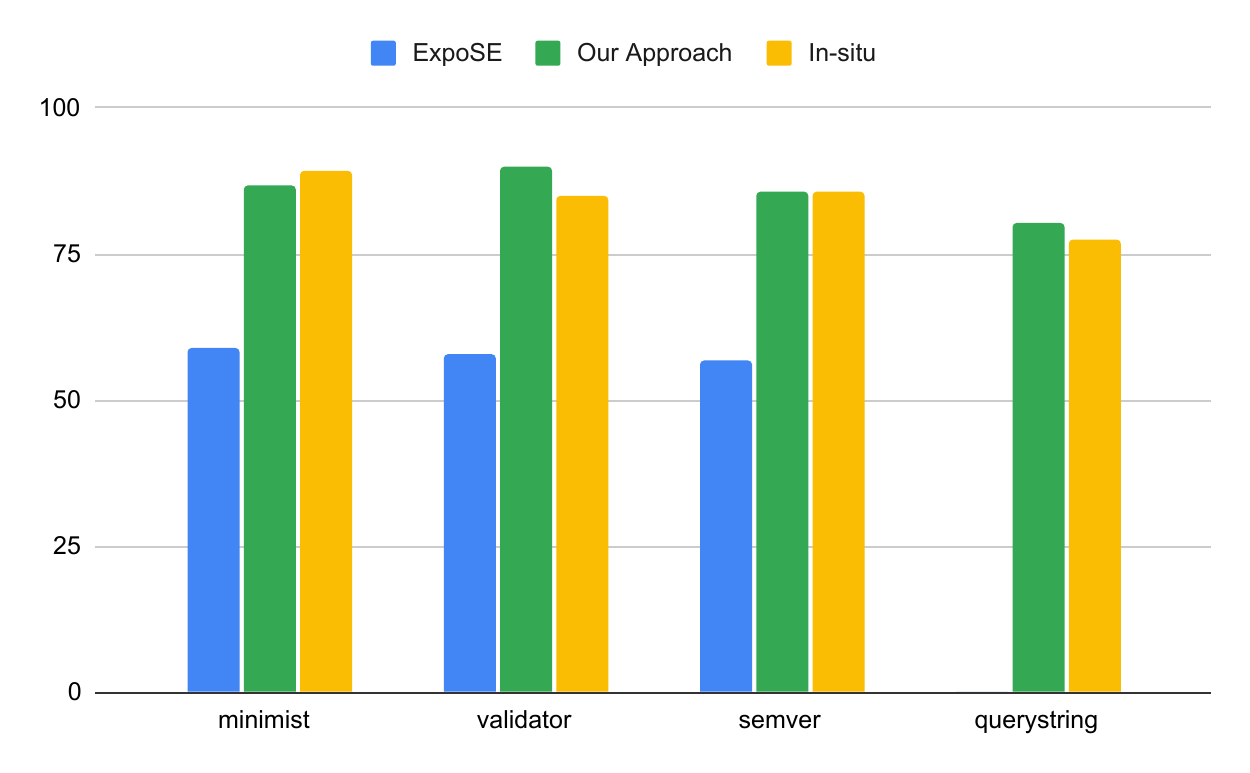}
\caption{Statement Coverage Comparison among our approach, In-situ approach and ExpoSE}
\label{fig:coveragecomparisontool}
\end{figure} 
\subsection{Bug Detection Efficiency}

As the test cases generated by our approach are replayed on the libraries under test, our method detected all the bugs that the in-situ method found. At the same time, our method only uses a fraction of the time that the in-situ approach needs. Typically, the in-situ approach takes about 3 to 5 minutes to complete an iteration of test case generation and it only needs about 5 seconds to complete an iteration with our approach. Exceptions are thrown during execution. There are two types of exceptions: unhandled and handled. The unhandled exceptions tend to indicate potential valuable bugs. The handled exceptions often indicate that the developers are aware of these exceptions, but want to deal with them later. Such exceptions are also valuable to both the developers and potential hackers, albeit less valuable than unhandled ones.

\begin{table*}
\centering
  \begin{tabular}{|l|l|l|}
    \hline
    \textbf{functions} & \textbf{Bugs} & \textbf{Known}\\
    \hline
    formatNumber & No boundary check for empty string & No \\
    \hline
    encodeDate & No NULL check for function argument & No \\ 
    \hline
    regexExec & Unhandled input syntax error& No \\
    \hline
    isVAT & Mishandled country code & No \\
    \hline
    chalkClass & Deprecated constructor invoked & Yes \\
    \hline
    stringify & Incorrect parsing of separators & Yes\\ 
    \hline
  \end{tabular}\\
  \caption{Bugs detected in functions}
  \label{tab:bugdetectedinfunction}
\end{table*}

Table~\ref{tab:bugdetected} shows a summary of the bugs that we discovered. The bug from \textit{benchmarkify} is a missing boundary check for empty string. It causes the \texttt{formatNumber} function to return a NULL object. When another function is later invoked on this NULL Object, it throws a TypeError exception. In the \texttt{encodeDate} function of \textit{msgpack5}, a parameter, \texttt{dt}, is used directly without checking for NULL value. In \textit{is-regex}, an input syntax error is not handled in the \texttt{regexExec} function. In \textit{validator}, a particular country code is not handled and leads to the execution of a \texttt{catch} block in the \texttt{isVAT} function.
In \textit{chalk}, a deprecated \texttt{constructor} is used in an \texttt{else} branch in the \texttt{chalkClass} function, causing an unhandled exception. In \textit{stringify}, incorrect parsing of separators in the \texttt{stringify} function causes an unhandled exception.

%% file: 06-relatedwork.tex
Our approach is closely aligned with prior research on execution tracing within native execution environments and enhancing trace translation in symbolic execution for JavaScript. Typically, the focus is on JavaScript scripts, including those running in browsers and browser-less runtimes like Node.js. Many symbolic execution techniques for JavaScript involve the development of application-specific symbolic execution engines or substantial modifications to JavaScript execution engines to facilitate symbolic execution. These methods often rely on intermediate representations during trace translation. For instance, SymJS is an example of a framework designed for testing client-side JavaScript scripts using symbolic execution~\cite{li2014symjs}. It modifies Rhino JS engine for symbolic execution~\cite{jin2008research}. 
For browser-less JavaScript, JALANGI is a framework for writing heavy-weight dynamic analysis, which can be enabled on JavaScript as a symbolic execution engine~\cite{sen2013jalangi}. COSETTE is another symbolic execution engine for JavaScript using an intermediate representation, namely JSIL, translated from JavaScript~\cite{santos2018symbolic}. ExpoSE applies symbolic execution on standalone JavaScript and uses JALANGI as its symbolic execution engine. ExpoSE's contribution is in addressing the limitation that JALANGI does not readily support regular expressions for JavaScript~\cite{loring2017expose}. Kudzu targeted AJAX applications by implementing a dynamic symbolic interpreter that takes a simplified intermediate language for JavaScript~\cite{kudzu}. To the best of our knowledge, no symbolic execution framework for JavaScript has directly utilized JavaScript's native execution environments for execution tracing~\cite{li2014two}.

%% file: 07-conclusion.tex
In this paper, our approach introduced improvements to the in-situ concolic testing of JavaScript. We have deployed a new execution tracer leveraging V8's Sparkplug baseline compiler to improve the tracing process and a new assembly to LLVM IR using \textit{remill} libraries. It improves the efficiency and effectiveness of the infrastructure of the in-situ concolic testing for JavaScript while keeping the native execution environments for JS scripts under test. We evaluated its effectiveness and efficiency by comparing the coverage, bug detection, and time consumption with the in-situ approach on the same test set, which are 160 Node.js libraries that heavily utilize the \texttt{String} type and its operations. The results show our approach achieve similar statement coverage on these libraries within no more than 10\% difference on average and is able to detect all bugs that are detected by the in-situ method, which only use a fraction of the time needed by the in-situ approach.